\documentclass[twocolumn,showpacs,preprintnumbers,prd,superscriptaddress,nofootinbib]{revtex4-1}

\usepackage{amsmath}
\usepackage{amsfonts}
\usepackage{amssymb}
\usepackage{graphicx}
\usepackage{color}
\usepackage{hyperref}
\usepackage{booktabs}
\usepackage{hyperref}
\usepackage{cleveref}
\usepackage{color}
\Crefrangeformat{equation}{Eqs. (#3#1#4)--(#5#2#6)}
\Crefname{equation}{Eq.}{Eqs.}
\Crefname{figure}{Fig.}{Figs.}
\Crefname{section}{Sec.}{Secs.}

\usepackage{etoolbox}
\makeatletter
\appto{\appendix}{%
  \@ifstar{\def\theequation@prefix{A.}}%
          {}%
}
\makeatother

\begin{document}

\title{Growth of matter perturbations in non-minimal teleparallel dark energy}

\author{Rocco D'Agostino}
\email{rocco.dagostino@roma2.infn.it}
\affiliation{Dipartimento di Fisica, Universit\`a degli Studi di Roma ``Tor Vergata'', Via della Ricerca Scientifica 1, I-00133, Roma, Italy.}
\affiliation{Istituto Nazionale di Fisica Nucleare (INFN), Sez. di Roma ``Tor Vergata'', Via della Ricerca Scientifica 1, I-00133, Roma, Italy.}

\author{Orlando Luongo}	
\email{orlando.luongo@lnf.infn.it}
\affiliation{Istituto Nazionale di Fisica Nucleare, Laboratori Nazionali di Frascati, 00044 Frascati, Italy.}
\affiliation{School of Science and Technology, University of Camerino, I-62032, Camerino, Italy.}
\affiliation{Instituto de Ciencias Nucleares, Universidad Nacional Aut\'onoma de M\'exico, AP 70543, M\'exico DF 04510, Mexico.}

\begin{abstract}
We study the growth rate of matter perturbations in the context of teleparallel dark energy in a flat universe. We investigate the dynamics of different theoretical scenarios based on specific forms of the scalar field potential. Allowing for non-minimal coupling between torsion scalar and scalar field, we perform a phase-space analysis of the autonomous systems of equations through the study of critical points. We thus analyze the stability of the critical points, and discuss the cosmological implications searching for possible attractor solutions at late times. Furthermore, combing the growth rate data and the Hubble rate measurements, we place observational constraints on the cosmological parameters of the models through Monte Carlo numerical method. We find that the scenario with a non-minimal coupling is favoured with respect to the standard quintessence case.
Adopting the best-fit results, we show that the dark energy equation of state parameter can cross the phantom divide. Finally, we compare our results with the predictions of the concordance $\Lambda$CDM paradigm by performing Bayesian model selection.
\end{abstract}

\maketitle


\section{Introduction}
\label{sec:intro}

Understanding the universe dynamics at all cosmic scales represents a challenge for theoretical cosmology \cite{copeland,otro}. In particular, observations show that the universe undergoes an accelerated phase at late times \cite{Haridasu17}. Baryons and cold dark matter cannot speed the universe up due to the action of gravity\footnote{This is consequence of the fact that matter is under the form of dust, i.e. provides a vanishing pressure. For alternative perspectives, see for example e.g. \cite{ioemarco,ioeherny}.}, so that one needs to include within Einstein's equation a bizarre fluid, dubbed dark energy, whose equation of state (EoS) is negative \cite{galaxy,miao}. Even though this scenario fairly well describes the large-scale dynamics, it fails to be predictive on dark energy's origin. Extended theories of gravity may represent a plausible landscape toward the determination of dark energy's nature as general relativity (GR) breaks down \cite{reviewcapozz}.

\noindent Although a wide number of modified gravity theories has been already investigated, we here assess modified teleparallel gravity models, initially proposed as alternatives to inflationary phases \cite{fT1}. In this picture, the way in which dark energy is recovered to speed up the universe today is provided by analytical functions of the torsion $T$, namely $f(T)$ \cite{fT2}. The approach of $f(T)$ candidates as a robust  alternative to barotropic dark energy fluids and to other extensions of Einstein's gravity \cite{altro1}.

\noindent Unfortunately, the function $f(T)$ is not known \emph{a priori}, so any suitable $f(T)$ models aim at describing the late-time dynamics might be either postulated or reconstructed through model-independent techniques \cite{fT3,fT4,fT5}.

We here circumscribe our attention to particular $f(T)$ models, which have reached great attention for their capability of describing both late and early epochs of universe's evolution \cite{altro2}. Specifically, we consider the teleparallel dark energy scenario with a non-minimal coupling between torsion scalar and scalar field \cite{Geng11,Xu12,Wei12}. We thus perform a phase-space analysis by studying the stability of critical points and search for late-times attractor solutions. In particular, if the real parts of the eigenvalues are negative, the corresponding critical point is stable and represents an attractor solution. We discuss the cosmological consequences of the theoretical models emerging from different choices of the scalar field potential. We analyze the case of null, constant, linear and exponential potentials.
Moreover, we study the growth rate of matter perturbations in terms of evolution of the density contrast. We also compare the observational constraints got from the latest available data with the standard outcomes of the $\Lambda$CDM model.

The paper is structured as follows.
In \Cref{sec:theory}, we review the main ingredients of teleparallel gravity and $f(T)$ cosmology.
In \Cref{sec:dynamics}, we describe the dynamical system approach to study the evolution of teleparallel dark energy with non-minimal coupling. We show how to obtain an autonomous system of equations by assuming suitable functional forms for the scalar field potential. Then, in \Cref{sec:phase-space} we perform a phase-space analysis of the dynamical equations and study the stability of critical points.
In \Cref{sec:growth}, we analyze the growth of matter overdensities and explain how the growth  factor data can be used to obtain predictions on our theoretical models.
In \Cref{sec:results}, we present numerical outcomes got from Monte Carlo analysis in terms of observational constraints over the cosmological parameters. Moreover, we compare our results with the predictions of the concordance $\Lambda$CDM paradigm, and we perform AIC and BIC analysis to select the best-performing model. Finally, \Cref{sec:conclusion} is dedicated to the discussion of the results and to the future perspectives of our work.


\section{$f(T)$ gravity and teleparallel dark energy}
\label{sec:theory}

In this section, we discuss the main features of modified teleparallel cosmology \cite{Bengochea09,Linder10,Capozziello15,D'Agostino17,Capozziello18,Abedi18}. Teleparallel gravity is described in terms of tetrad fields $e_A(x^\mu)$, forming an orthonormal basis for the tangent space of the manifold at each point $x^\mu$\footnote{In our notation, capital Latin indices run over the tangent space, while Greek indices run over the manifold.}. In this representation, the tangent space makes use of the Minkowski metric $\eta_{AB}=\text{diag}(1,-1,-1,-1)$, whereas the metric tensor is given in terms of dual vierbeins $e^A(x^\mu)$ by:
\begin{equation}
g_{\mu\nu}=\eta_{AB}\ e^A_\mu e^B_\nu\ .
\end{equation}
In lieu of Levi-Civita connections used in GR, teleparallel gravity uses the Weitzenb\"{o}ch connections $\hat{\Gamma}_{\mu\nu}^{\lambda}$, characterized by vanishing curvature with non-zero torsion. The torsion scalar is thus \cite{Cai16}:
\begin{equation}
T={S_\rho}^{\mu\nu}{T^{\rho}}_{\mu\nu}\ ,
\end{equation}
where
\begin{equation}
{S_{\rho}}^{\mu\nu}\equiv\dfrac{1}{2}\left({K^{\mu\nu}}_\rho+\delta^\mu_\rho\ {T^{\alpha\nu}}_\alpha- \delta_\rho^\nu\ {T^{\alpha\mu}}_\alpha \right),
\end{equation}
and $T_{\mu\nu}$ and ${K^{\mu\nu}}_\rho$ are the torsion tensor and the contorsion tensor, respectively defined as
\begin{align}
&T_{\mu\nu}^\lambda\equiv \hat{\Gamma}_{\mu\nu}^{\lambda}-\hat{\Gamma}_{\nu\mu}^{\lambda}=e^\lambda_A(\partial_\mu e_\nu^A-\partial_\nu e_\mu^A)\ , \\
&{K^{\mu\nu}}_\rho	\equiv-\dfrac{1}{2}\left({T^{\mu\nu}}_\rho-{T^{\nu\mu}}_\rho-{T_{\rho}}^{\mu\nu}\right) .
\end{align}
The simplest action of teleparallel gravity can be written as $\mathcal{S}=\int d^4 x\ e \, [T/(2\kappa)+\mathcal{L}_m]$,
where $e=\det(e^A_\mu)=\sqrt{-g}$, $\kappa=8\pi G$ and $\mathcal{L}_m$ is the Lagrangian density for matter  \cite{Hayashi79,Arcos04}.
As one varies the above action  with respect to the vierbeins, the corresponding field equations are equivalent to the Einstein ones, leading to the well-known  \emph{Teleparallel Equivalent to General Relativity} (TEGR).

A first generalization of TEGR is inspired by the $f(R)$ gravity theories. In fact, one can substitute $T$  in $\mathcal{S}$ with a generic function of the torsion scalar, defining the so-called $f(T)$ theories \cite{Bengochea11,Ferraro07,Dent11}.
Another landscape is the addition of a canonical scalar field, reproducing the \textit{quintessence} effects. In this respect, one can modify the action $\mathcal{S}$ to include a non-minimal coupling between $T$ and a scalar field $\phi$:
\begin{equation}
\mathcal{S}=\int d^4x\ e \left[\dfrac{T}{2\kappa}+\dfrac{1}{2}\left(\partial_\mu\phi\partial^\mu \phi+\xi T\phi^2\right)-V(\phi)+\mathcal{L}_m\right],
\label{action}
\end{equation}
where $V(\phi)$ is the scalar field potential, and $\xi$ is the coupling constant. The non-minimal quintessence\footnote{In the standard scenario, the coupling is between the Ricci scalar and the scalar field.} in the framework of teleparallel gravity is named \emph{teleparallel dark energy} \cite{Geng11,Kucukakca}. Here, varying the action (\ref{action}) with respect to the vierbeins provides the field equations:
\begin{align}
&\left[\dfrac{1}{e}\partial_\mu(ee_A^\rho {S_\rho}^{\mu\nu})-e_A^\lambda {T^\rho}_{\mu\lambda}{S_\rho}^{\nu\mu}-\dfrac{1}{4}e^\nu_AT\right]\left(\dfrac{2}{\kappa}+2\xi\phi^2\right) \nonumber \\
&+e_A^\mu\partial^\nu\phi\partial_\mu\phi+4\xi e_A^\rho {S_\rho}^{\mu\nu}\phi \ \partial_\mu\phi-e_A^\nu\left[\dfrac{1}{2}\partial_\mu\phi\partial^\mu\phi-V(\phi)\right]  \nonumber \\
&=e_A^\rho  {{T^{(m)}}_\rho}^\nu\ ,
\end{align}
where ${{T^{(m)}}_\rho}^\nu$ is the matter energy-momentum tensor.

\subsection{Cosmology of $f(T)$ gravity}
We search for cosmological solutions by considering the spatially flat Friedmann-Lema\^{i}tre-Robertson-Walker metric\footnote{Throughout the text, we use units such that $c=1$.} $ds^2=dt^2-a(t)^2\delta_{ij}dx^idx^j$,
where $a(t)$ is the cosmic scale factor. In this case, the vierbein fields read $e_A^\mu=\text{diag}(1,a,a,a)$,  having $T=-6H^2$, where $H\equiv\dot{a}/a$ is the Hubble parameter.  The modified Friedmann equations with  a perfect fluid source are:
\begin{align}
&H^2=\dfrac{\kappa}{3}\left(\rho_m+\rho_\phi\right), \label{eq:Friedmann1}\\
&\dot{H}=-\dfrac{\kappa}{2}\left(\rho_m+\rho_\phi+p_\phi\right), \label{eq:Friedmann2}
\end{align}
where we have neglected the contribution of radiation, assuming pressureless matter. Here, $\rho_m$ is the matter energy density obeying the continuity equation $\dot{\rho}_m+3H\rho_m=0$,
while $\rho_\phi$ and $p_\phi$ represent the energy density and pressure of the scalar field, respectively given by
\begin{align}
&\rho_\phi=\dfrac{1}{2}\dot{\phi}^2+V(\phi)-3H^2\xi\phi^2\ , \label{eq:rho_phi}\\
&p_\phi=\dfrac{1}{2}\dot{\phi}^2-V(\phi) +4H\xi\phi\dot{\phi}+\left(3H^2+2\dot{H}\right)\xi\phi^2 \label{eq:p_phi}\ .
\end{align}
The Klein-Gordon equation is obtained by varying the action (\ref{action}) with respect to the scalar field $\phi$:
\begin{equation}
\ddot{\phi}+3H\dot{\phi} +V_{,\phi}=\xi T \phi\ ,
\label{eq:Klein-Gordon}
\end{equation}
where $V_{,\phi}\equiv dV/d\phi$ and $\xi T \phi$ represents the source term, which provides  the standard Klein-Gordon without source as $\xi\rightarrow0$. In this limit, teleparallel dark energy reduces to ordinary quintessence at both background and perturbation levels. Further, in such a scenario the scalar field plays the role of dark energy, so that \Cref{eq:Klein-Gordon} is equivalent to the conservation equation
\begin{equation}
\dot{\rho}_\phi+3H(1+w_\phi)\rho_\phi=0\ ,
\end{equation}
where $w_\phi\equiv p_\phi/\rho_\phi$ coincides with the dark energy EoS that can cross the phantom divide, as one can clearly see from \Cref{eq:rho_phi,eq:p_phi}.


\section{Dynamical system of equations}
\label{sec:dynamics}

A powerful and elegant way to investigate the universe dynamics is to recast the cosmological equations into a dynamical system. This approach permits to combine analytical and numerical methods to obtain quantitative information on the models under study.
One can examine the asymptotic states corresponding to different phases of the cosmic evolution by studying the invariant manifolds for which one equation of the dynamical system vanishes \cite{Hrycyna17}, or by searching for the critical points that nullify the derivatives of the dynamical variables. The latter approach is the one we pursue in the present work.
To that purpose, we introduce the following dimensionless variables:
\begin{subequations}
\begin{align}
&x\equiv \dfrac{\kappa\dot{\phi}}{\sqrt{6}H}\ ,  \hspace{0.8cm}y\equiv \dfrac{\kappa\sqrt{V}}{\sqrt{3}H}\ ,\\
&u\equiv \kappa\phi \ , \hspace{1.25cm} v\equiv \dfrac{\kappa \sqrt{\rho_m}}{\sqrt{3}H}\ .
\end{align}
\end{subequations}
Then, \Cref{eq:Friedmann1} can be recast under the form
\begin{equation}
x^2+y^2-\xi u^2+v^2=1\ .
\label{condition}
\end{equation}
Moreover, from \Cref{eq:Friedmann2,eq:rho_phi,eq:p_phi} one finds
\begin{equation}
s\equiv-\dfrac{\dot{H}}{H^2}=\dfrac{3x^2+2\sqrt{6}\ \xi x u +3v^2/2}{1+\xi u^2}\ .
\end{equation}
We can thus rewrite the conservation equations for $\rho_m$ and $\rho_\phi$ as a dynamical system for the new variables:
\begin{equation}
\left\{
\begin{aligned}
&x'= (s-3)x-\sqrt{6}\ \xi u-\dfrac{\kappa V_{,\phi}}{\sqrt{6}H^2}\ , \\
&y'= sy+\dfrac{x}{\sqrt{2}H}\dfrac{V_{,\phi}}{\sqrt{V}}\ , \\
&u'= \sqrt{6}\ x \ .
\end{aligned}
\right .
\label{system}
\end{equation}
We note that the evolution of $v$ is given in terms of the other dynamical variables as follows from \Cref{condition}.
The `prime' here denotes the derivative with respect to the number of $e$-folds, $N\equiv \ln a$. In this formalism, the energy densities of the cosmic species read
\begin{equation}
\Omega_m=v^2\ , \hspace{1cm} \Omega_\phi=x^2+y^2-\xi u^2\ .
\label{eq:Omega_m and Omega_phi}
\end{equation}
The system of differential equations given by Eqs.~(\ref{system}) can be numerically solved by choosing appropriate initial values for the variables $\{y,u,v\}$. Following \cite{Geng13}, we choose $y_{init}=10^{-6}$, $u_{init}=10^{-6}$ and $v_{init}^2=0.999$ at $a_{init}=10^{-2}$. The solutions turn out to be stable for $y_{init}\leq10^{-5}$, $u_{init}\in[10^{-9},10^{-5}]$, $v_{init}^2\geq 0.999$, and $a_{init}\in[10^{-2},10^{-1}]$. Outside those ranges, the numerical procedure exhibits convergence issues.
The dark energy EoS is given by
\begin{equation}
w_\phi=\dfrac{x^2-y^2+\xi u^2-\frac{2}{3}\ \xi s u^2+4\sqrt{\frac{2}{3}}\ \xi x u}{x^2+y^2-\xi u^2}\ .
\label{eq:w}
\end{equation}
Once a suitable form of $V(\phi)$ is chosen, Eqs.~(\ref{system}) becomes an autonomous system and one can determine the dynamics of the cosmological equations.
In the following, we consider three specific forms of scalar field potential, motivating each of them through physical reasons.

\subsection{Vanishing potential}

\noindent The first model makes use of a vanishing potential:
\begin{equation}
V(\phi)=0\ .
\label{V=0}
\end{equation}
This represents the simplest scenario, characterized by a tracker behaviour at early times. In this model, the late-time acceleration is realized for $\xi<0$ \cite{Gu13}.
As \Cref{V=0} holds, the dynamical system simply reduces to
\begin{equation}
\left\{
\begin{aligned}
&x'= (s-3)x-\sqrt{6}\ \xi u\ , \\
&u'= \sqrt{6}\ x \ .
\end{aligned}
\right .
\label{system V=0}
\end{equation}
This model describes a field $\phi$, free from interactions and may be used as prototype to characterize non-bounded phions.

\subsection{Constant potential}

\noindent The third case is a teleparallel dark energy model with constant potential function:
\begin{equation}
V(\phi)=V_0\ .
\end{equation}
This type of potential has been considered in \cite{Hrycyna17}, where the authors showed that the dynamical system describing the evolution of a flat FLRW model with a non-minimally coupled scalar field is equipped with an invariant de Sitter manifold.
In the case of a constant potential, system (\ref{system}) takes the simple form:
\begin{equation}
\left\{
\begin{aligned}
&x'= (s-3)x-\sqrt{6}\ \xi u\ , \\
&y'= sy \ , \\
&u'= \sqrt{6}\ x \ .
\end{aligned}
\right .
\label{system V=const}
\end{equation}

\subsection{Linear potential}

\noindent The second model is teleparallel dark energy with a linear potential:
\begin{equation}
V(\phi)=V_0\kappa \phi\ .
\label{V linear}
\end{equation}
It has been shown that a linear potential provides a possible solution to the coincidence problem \cite{Avelino05}. In addition, this form seems to be favoured even by the anthropic principle, i.e.  galaxy formation is possible \emph{only} in regions where $V(\phi)$ is well approximated by a linear function \cite{Garriga}.
In view of \Cref{V linear}, system (\ref{system}) reads
\begin{equation}
\left\{
\begin{aligned}
&x'= (s-3)x-\sqrt{6}\ \xi u-\sqrt{\dfrac{3}{2}}\dfrac{y^2}{u}\ , \\
&y'= sy+\sqrt{\dfrac{3}{2}} \dfrac{xy}{u} \ , \\
&u'= \sqrt{6}\ x \ .
\end{aligned}
\right .
\label{system V linear}
\end{equation}

\subsection{Exponential potential}

\noindent Finally, we consider an exponential potential:
\begin{equation}
V(\phi)=V_0 e^{-\kappa \phi}\ .
\label{V exp}
\end{equation}
Exponential potentials have been widely used in the literature to study inflation in the early universe, structure formation and dark energy dynamics \cite{Ferreira97,Copeland98}.
For a scalar field potential of the form (\ref{V exp}), the dynamical system becomes
\begin{equation}
\left\{
\begin{aligned}
&x'= (s-3)x-\sqrt{6}\ \xi u-\sqrt{\dfrac{3}{2}}y^2\ , \\
&y'= sy-\sqrt{\dfrac{3}{2}} xy \ , \\
&u'= \sqrt{6}\ x \ .
\end{aligned}
\right .
\label{system V exp}
\end{equation}

\section{Phase-space analysis}
\label{sec:phase-space}

In this section we perform a phase-space analysis for the autonomous systems obtained above. Since in the case of null potential no critical points have been found, we focus on the non-vanishing potentials. To do so, we introduce the vectors $\mathbf{X}=(x,y,u)$ and $\mathbf{X'}=(x',y',u')$.
We thus find the critical points $\mathbf{X}_c$ satisfying the equation $\mathbf{X}'=0$. To study the stability of the critical points, we set $\mathbf{X}=\mathbf{X}_c+\mathbf{\delta X}$, where $\mathbf{\delta X}=(\delta x, \delta y, \delta u)$ are linear perturbations of the dynamical variables. Then, we linearize the dynamical equations to get $\delta \mathbf{X}'=\mathcal{M}\ \delta\mathbf{X}$, where $\mathcal{M}$ is the coefficients matrix. Finally, the eigenvalues of $\mathcal{M}$ evaluated at each critical point determine its stability. In particular, if the real parts of all the eigenvalues are negative, the corresponding critical point is stable and represents an attractor solution at late times.

\subsection{The case $V(\phi)=V_0$}

In the case of constant potential, $\mathbf{X}'=0$ yields the following critical point:
\begin{equation}
\mathbf{X}_c=(0,1,0)\ .
 \label{crit const}
\end{equation}
The cosmological behaviour in correspondence of the critical point is obtained from \Cref{eq:w,eq:Omega_m and Omega_phi}, which give
\begin{equation}
\begin{aligned}
&w_{\phi}=-1\ , \\
&\Omega_{m}=0\ , \\
&\Omega_{\phi}=1 \ .
\end{aligned}
\end{equation}
This scenario corresponds to a de Sitter solution, in which the universe is dominated by dark energy in the form of a cosmological constant.

\begin{figure}[h]
\begin{center}
\includegraphics[width=3.2in]{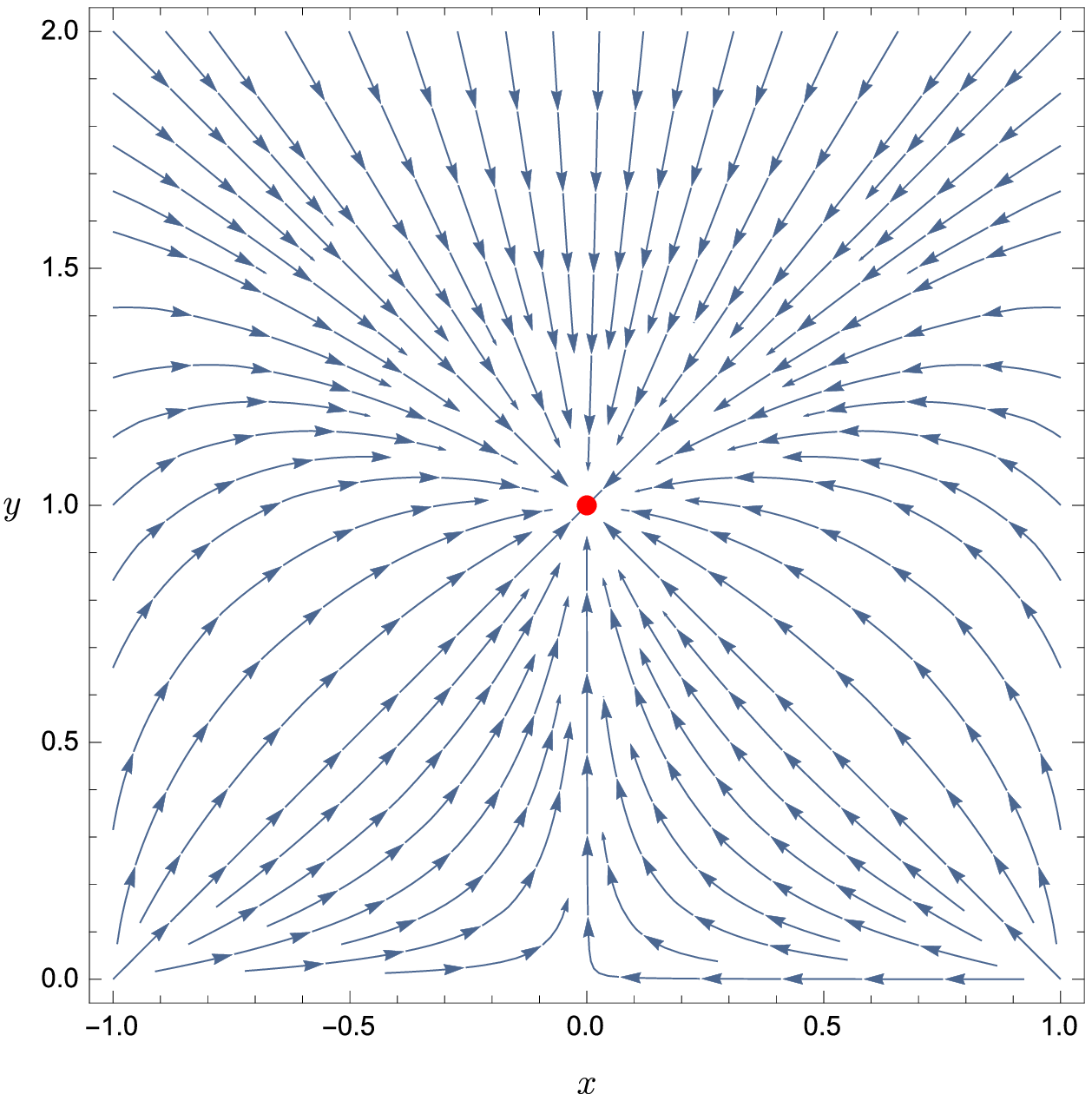}
\caption{Phase-space trajectories on the $x$ - $y$ plane for teleparallel dark energy with $V(\phi)=V_0$ and $\xi=1$. The red dot corresponds to the critical point  (\ref{crit const}), which represents an attractor solution of the dynamical system.}
\label{fig:attractor1}
\end{center}
\end{figure}

We study the stability of the critical point through linear perturbations of the dynamical equations:
\begin{align}
\delta x'&=(s-3)\delta x +x \delta s-\sqrt{6}\xi \delta u , \\
\delta y'&=s\delta y+y\delta s \ , \\
\delta u'&=\sqrt{6}\ \delta x\ ,
\end{align}
where
\begin{align}
s=&\ \dfrac{3x^2+2\sqrt{6} \xi x u +3(1-x^2-y^2+\xi u^2)/2}{1+\xi u^2}\ , \label{eq:s} \\
\delta s=&\ \dfrac{6x\delta x+2\sqrt{6}\xi(x\delta u+u\delta x)-3(x\delta x+y\delta y-\xi u \delta u)}{1+\xi u^2} \nonumber \\
&\ -\dfrac{2\xi u \delta u}{1+\xi u^2}s\ . \label{eq:delta s}
\end{align}
The coefficients of the perturbation matrix are reported in \Cref{app:matrix const}. The eigenvalues at the critical point are found by solving the equation
\begin{equation}
(3+\mu)(\mu^2+3\mu+6\xi)=0\ ,
\end{equation}
whose solutions are:
\begin{align}
&\mu_1=-3\ , \\
&\mu_2=\dfrac{1}{2}\left(-3-\sqrt{9-24\xi}\right) , \\
&\mu_3=\dfrac{3}{2}\left(-3+\sqrt{9-24\xi}\right) .
\end{align}
The real parts of the above eigenvalues are all negative for $\xi>0$. Thus, under the condition $\xi>0$, the critical point is stable and it corresponds to an attractor solution for the universe at late times (see \Cref{fig:attractor1}).

\subsection{The case $V(\phi)=V_0\kappa\phi$}
In the case of linear potential,  $\mathbf{X}'=0$ together with \Cref{condition} and the conditions $y\geq 0$ provide the following critical points:
\begin{align}
&\mathbf{X}^{(1)}_c=\bigg(0,\ \sqrt{\dfrac{2}{3}},\ \dfrac{-1}{\sqrt{3}\sqrt{-\xi}}\bigg) , \label{crit1 lin} \\
&\mathbf{X}^{(2)}_c=\bigg(0,\ \sqrt{\dfrac{2}{3}},\ \dfrac{1}{\sqrt{3}\sqrt{-\xi}} \bigg) , \label{crit2 lin}
\end{align}
which exist if $\xi<0$.
The cosmological implications are thus obtained by evaluating \Cref{eq:w,eq:Omega_m and Omega_phi} at points $\mathbf{X}^{(1)}_c$ and $\mathbf{X}^{(2)}_c$ . In both cases, we find
\begin{equation}
\begin{aligned}
&w_{\phi}=-1\ , \\
&\Omega_{m}=0\ , \\
&\Omega_{\phi}=1 \ .
\end{aligned}
\end{equation}
This set represents a de Sitter universe, completely dominated by dark energy.

To study the stability of the critical points, we write down the evolution equations for the linear perturbations:
\begin{align}
\delta x'&=(s-3)\delta x +x \delta s-\sqrt{6}\xi \delta u-\sqrt{\dfrac{3}{2}}\dfrac{y}{u}\left(2\delta y-yu^{-1} \delta u\right) , \\
\delta y'&=s\delta y+y\delta s +\sqrt{\dfrac{3}{2}}u^{-1}\left(x\delta y+y\delta x-x y u^{-1}\delta u\right) , \\
\delta u'&=\sqrt{6}\ \delta x\ ,
\end{align}
where $s$ and $\delta s$ are given by \Cref{eq:s,eq:delta s}, respectively.
We report the coefficients matrix of the perturbation equations in \Cref{app:matrix lin}. The critical points (\ref{crit1 lin}) and (\ref{crit2 lin}) give the same eigenvalues, which are obtained by solving the equation
\begin{equation}
(3+\mu)(\mu^2+3\mu+18\xi)=0\ .
\end{equation}
One thus gets
\begin{align}
&\mu_1=-3\ , \\
&\mu_2=-\dfrac{3}{2}\left(1+\sqrt{1-8\xi}\right) , \\
&\mu_3=-\dfrac{3}{2}\left(1-\sqrt{1-8\xi}\right) .
\end{align}
The real part of $\mu_3$ is negative only if $\xi>0$, which is against the existence condition $\xi<0$ for the critical points. Therefore, the critical points are unstable and no attractor solutions exist for this model.

\subsection{The case $V(\phi)=V_0e^{-\kappa\phi}$}

In this case, imposing $\mathbf{X}'=0$ and the condition $y\geq 0$ give three critical points:
\begin{align}
&\mathbf{X}^{(1)}_c=\left(0,\ 0, \ 0 \right) , \label{crit1 exp} \\
&\mathbf{X}^{(2)}_c=\bigg(0,\ \sqrt{2\xi -2\sqrt{\xi(\xi-1)}},\ 1-\sqrt{\dfrac{\xi-1}{\xi}}\bigg) ,  \label{crit2 exp} \\
&\mathbf{X}^{(3)}_c=\bigg(0,\ \sqrt{2\xi +2\sqrt{\xi(\xi-1)}},\ 1+\sqrt{\dfrac{\xi-1}{\xi}} \bigg).  \label{crit3 exp}
\end{align}
We note that point $\mathbf{X}^{(1)}_c$ always exists, whereas point $\mathbf{X}^{(2)}_c$ exists if $\xi\geq 1$, and $\mathbf{X}^{(3)}_c$ exists if $\xi<0$ or $\xi\geq1$.
The cosmological behaviours at the critical points are found from \Cref{eq:w,eq:Omega_m and Omega_phi}. In particular, at point $\mathbf{X}^{(1)}_c$ one has
\begin{equation}
\begin{aligned}
&\Omega_{m}=1\ , \\
&\Omega_{\phi}=0 \ .
\end{aligned}
\end{equation}
This set corresponds to an non-accelerating Einstein-de Sitter universe, completely dominated by matter. On the other hand, at both points $\mathbf{X}^{(2)}_c$ and $\mathbf{X}^{(3)}_c$ one finds a de Sitter solution:
\begin{equation}
\begin{aligned}
&w_{\phi}=-1\ , \\
&\Omega_{m}=0\ , \\
&\Omega_{\phi}=1 \ .
\end{aligned}
\end{equation}
We study the stability of the critical points from the evolution equations for linear perturbations. In this case, we have
\begin{align}
\delta x'&=(s-3)\delta x +x \delta s-\sqrt{6}\xi \delta u+\sqrt{6}y \delta y  , \\
\delta y'&=s\delta y+y\delta s +\sqrt{\dfrac{3}{2}}\left(x\delta y+y\delta x\right) , \\
\delta u'&=\sqrt{6}\ \delta x\ ,
\end{align}
where $s$ and $\delta s$ are given as in \Cref{eq:s,eq:delta s}, respectively. The matrix with the coefficients of the perturbation equations is reported in \Cref{app:matrix exp}.
The eigenvalues $\mu_i^{(1)}$ for point $\mathbf{X}^{(1)}_c$ are obtained as solutions of the equation
\begin{equation}
(3 - 2 \mu)(2\mu^2+3\mu+12\xi)=0\ .
\end{equation}
We thus find
\begin{align}
&\mu_1^{(1)}=\dfrac{3}{2}\ , \\
&\mu_2^{(1)}= \dfrac{1}{4}\left(-3 - \sqrt{9 - 96 \xi}\right) , \\
&\mu_3^{(1)}=\dfrac{1}{4}\left(-3 + \sqrt{9 - 96 \xi}\right) .
\end{align}
Since $\mu_1^{(1)}$ is always positive, the critical point $\mathbf{X}^{(1)}_c$ is unstable.

\begin{figure}[h]
\begin{center}
\includegraphics[width=3.2in]{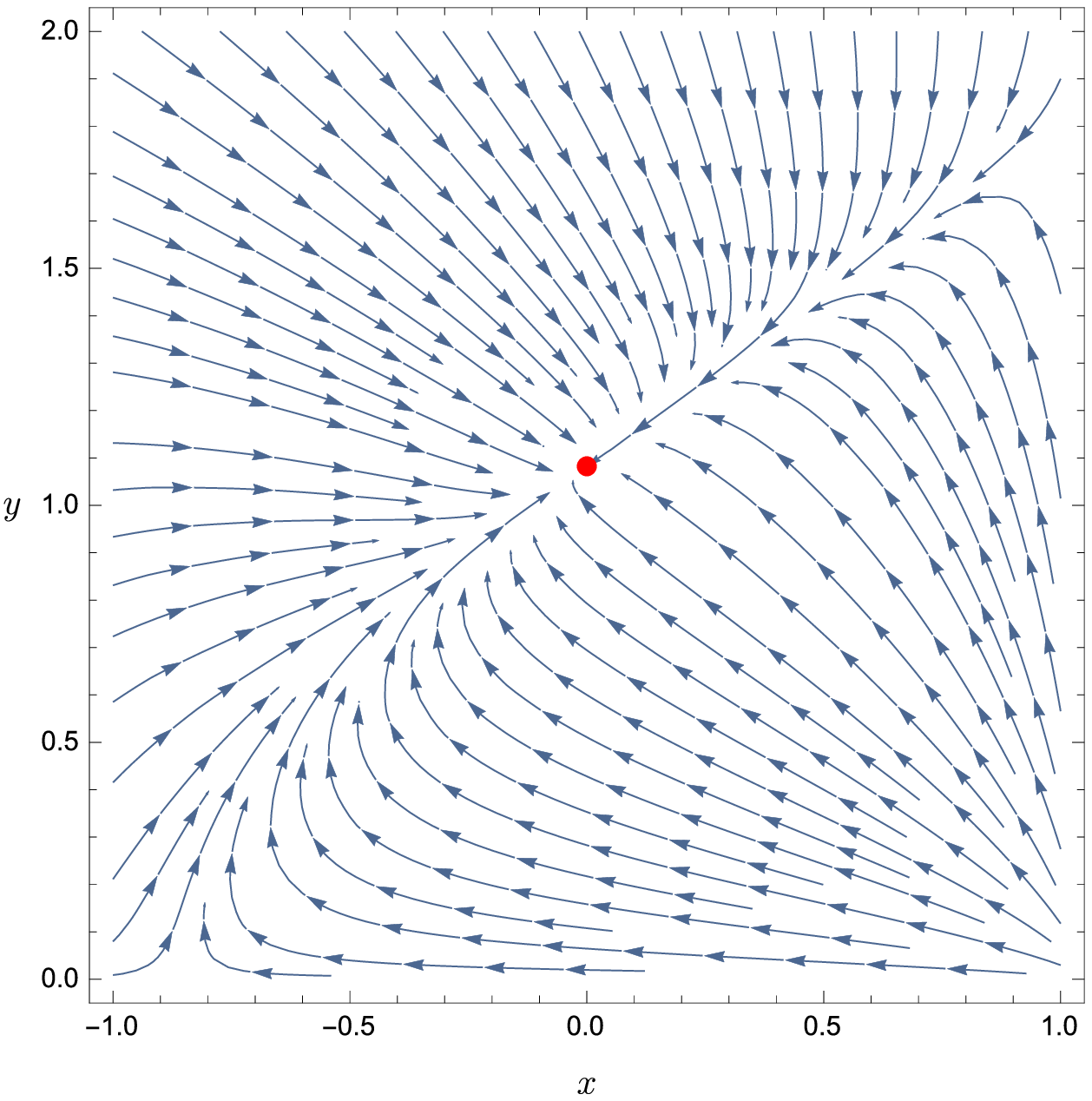}
\caption{Phase-space trajectories on the $x$ - $y$ plane for teleparallel dark energy with $V(\phi)=V_0e^{-\kappa\phi}$ and $\xi=2$. The red dot corresponds to the critical point $\mathbf{X}^{(2)}_c$ (cf. \Cref{crit2 exp}), which represents an attractor solution of the dynamical system.}
\label{fig:attractor}
\end{center}
\end{figure}

The eigenvalues $\mu_i^{(2)}$ corresponding to point  $\mathbf{X}^{(2)}_c$ are obtained from
\begin{widetext}
\begin{equation}
\dfrac{\xi(3 + \mu)\left[3 \mu + \mu^2 +  6 \left(-2 \xi + \sqrt{\xi(\xi-1)}\right) - 2 \left(-3 \mu - \mu^2 + 6 \xi\right) \left(-\xi + \sqrt{\xi(\xi-1)}\right)\right]}{\left(\xi- \sqrt{\xi(\xi -1)}\right)^2}=0\ ,
\end{equation}
\text{which gives}
\begin{align}
&\mu_1^{(2)}=-3\ , \\
&\mu_2^{(2)}=-\dfrac{3 - 6 \xi + 6 \sqrt{\xi(\xi-1)} + \sqrt{3} \left[3 + 64 \xi^3 + 4 \sqrt{\xi(\xi-1)} +  8 \xi \left(1 + 5 \sqrt{\xi(\xi-1)}\right) - 8 \xi^2 \left(9 + 8 \sqrt{\xi(\xi-1)}\right)\right]^{1/2}}{2 - 4 \xi + 4 \sqrt{\xi(\xi-1)}}\ , \\
&\mu_3^{(2)}=\dfrac{-3 + 6 \xi - 6 \sqrt{\xi(\xi-1)} + \sqrt{3} \left[3 + 64 \xi^3 + 4 \sqrt{\xi(\xi-1)} +  8 \xi \left(1 + 5 \sqrt{\xi(\xi-1)}\right) - 8 \xi^2 \left(9 + 8 \sqrt{\xi(\xi-1)}\right)\right]^{1/2}}{2 - 4 \xi + 4 \sqrt{\xi(\xi-1)}}\ .
\end{align}
\end{widetext}
The real parts of $\mu_i^{(2)}$ are all negative for $\xi>1$, where the critical points exist. Hence, we conclude that point $\mathbf{X}^{(2)}_c$ is stable and it represents an attractor for the universe at late times (see \Cref{fig:attractor}).

Finally, the eigenvalues $\mu_i^{(3)}$ of point $\mathbf{X}^{(3)}_c$ are found by solving
\begin{widetext}
\begin{equation}
\dfrac{\xi(3 + \mu)\left[-3 \mu - \mu^2 +  6 \left(2 \xi + \sqrt{\xi(\xi-1)}\right) - 2 \left(-3 \mu - \mu^2 + 6 \xi\right) \left(-\xi + \sqrt{\xi(\xi-1)}\right)\right]}{\left(\xi+ \sqrt{\xi(\xi -1)}\right)^2}=0\ ,
\end{equation}
\text{which provides}
\begin{align}
&\mu_1^{(3)}=-3\ , \\
&\mu_2^{(3)}=-\dfrac{-3 + 6 \xi + 6 \sqrt{\xi(\xi-1)} + \sqrt{3} \left[3 + 64 \xi^3 - 4 \sqrt{\xi(\xi-1)} +  8 \xi \left(1 - 5 \sqrt{\xi(\xi-1)}\right) + 8 \xi^2 \left(-9 + 8 \sqrt{\xi(\xi-1)}\right)\right]^{1/2}}{-2 + 4 \xi + 4 \sqrt{\xi(\xi-1)}}\ , \\
&\mu_3^{(3)}=\dfrac{3 - 6 \xi - 6 \sqrt{\xi(\xi-1)} + \sqrt{3} \left[3 + 64 \xi^3 - 4 \sqrt{\xi(\xi-1)} +  8 \xi (1 - 5 \sqrt{\xi(\xi-1)}) + 8 \xi^2 (-9 + 8 \sqrt{\xi(\xi-1)})\right]^{1/2}}{-2 + 4 \xi + 4 \sqrt{\xi(\xi-1)}}\ .
\end{align}
\end{widetext}
It turns out that the real parts of $\mu_i^{(3)}$ are not all negative for $\xi<0$ and $\xi \geq 1$. This implies that point $\mathbf{X}^{(3)}_c$ is unstable and no attractor solution corresponds to this critical point.

\section{Growth rate of matter perturbations}
\label{sec:growth}

Over the past years, it has become clear that the study of the background cosmological dynamics is not sufficient to discriminate between modified theories and the standard cosmological model. The study of density perturbations over the homogeneous and isotropic background could be a possible way to test possible deviations from GR and to break the degeneracy among the alternative theories \cite{degeneracy}. Indeed, any modification of GR would affect also the growth of cosmological perturbations. An important probe in this respect is the evolution of linear matter density contrast $\delta_m\equiv \delta\rho_m/\rho_m$ \cite{Tsujikawa07,DeFelice10}:
\begin{equation}
\dfrac{d^2\delta_m}{da^2}+\left(\dfrac{3}{a}+\dfrac{1}{E}\dfrac{dE}{da}\right)\dfrac{d\delta_m}{da}-\dfrac{3}{2}\dfrac{\Omega_{m0}}{a^5E^2}\dfrac{G_{eff}}{G}\delta_m=0\ ,
\label{eq:perturbations}
\end{equation}
where $E(a)\equiv H(a)/H_0$ and the subscript `0' refers to quantities evaluated at the present time.
In the case of a flat universe with an evolving dark energy EoS, the dynamics of the cosmic expansion is obtained as
\begin{equation}
E^2(a)=\Omega_{m0}a^{-3}+\Omega_{\phi0}\exp\left\{-3\int_a^1\left( \dfrac{1+w_{\phi}}{a'}\right)da'\right\} ,
\label{eq:E(a)}
\end{equation}
where $\Omega_{\phi0}=1-\Omega_{m0}$.
In the quasi static approximation and in the sub-horizon regime, the effective gravitational constant $G_{eff}$ is given by \cite{Abedi18}
\begin{equation}
G_{eff}=\dfrac{G}{1+\kappa \xi \phi^2}\left(1-\dfrac{\kappa\dot{\phi}^2}{2H^2(1+\kappa\xi\phi^2)}\right) .
\label{eq:G_eff}
\end{equation}
We note that in the case of minimal quintessence $(\xi=0$), the kinetic energy of the scalar field is subdominant with respect to the dark energy at the present time, so that one recovers the expected result $G_{eff}\simeq G$.
In terms of the dynamical variables $x$ and $u$, \Cref{eq:G_eff} becomes
\begin{equation}
\dfrac{G_{eff}}{G}=\dfrac{1}{1+\xi u^2}\left(1-\dfrac{3x^2}{1+\xi u^2}\right) .
\end{equation}

To study the growth rate of matter density perturbations, we introduce the quantity \cite{Nesseris17}
\begin{equation}
f(a)\equiv \dfrac{d\delta_m}{d \ln a}\ .
\label{eq:f(a)}
\end{equation}
Measurements from redshift space distortion and weak lensing have been obtained in the redshift interval $0<z<2$ for the factor
\begin{equation}
f\sigma_8(z)\equiv f(z)\sigma_8(z)\ ,
\end{equation}
where $\sigma_8(a)=\sigma_8 \delta_m(a)/\delta_m(1)$ is the rms fluctuations of the linear density field inside a radius of $8h^{-1}$Mpc, and $\sigma_8$ is its present day value.


\section{Observational constraints}
\label{sec:results}

We studied the observational viability of the non-minimal teleparallel scenarios described above by placing constraints on the present  matter density $\Omega_{m0}$, the present matter fluctuation amplitude $ \sigma_8$ and the coupling constant $\xi$. To this end, we used the ``Gold-2017" compilation of 18  independent $f\sigma_8$ measurements presented in \cite{Nesseris17}.
The model-dependence of these measurements required a correction by the assumed fiducial cosmology. We thus defined the ratio
\begin{equation}
\rho(z)=\dfrac{H(z)d_A(z)}{H_{fid}(z)d_{A,fid}(z)}\ ,
\label{correction}
\end{equation}
where $d_A(z)$ is the angular diameter distance given by
\begin{equation}
d_A(z)=(1+z)^{-1}\int_0^z\dfrac{dz'}{H(z')}\	.
\end{equation}
In the case of the data sample considered here, the fiducial model refers to the $\Lambda$CDM model, for which the Hubble expansion reads
\begin{equation}
H_{fid}(z)=H_0\sqrt{\Omega_{m0}(1+z)^3+\Omega_{\Lambda 0}}\ ,
\end{equation}
where $\Omega_{\Lambda 0}=1-\Omega_{m0}$.
Hence, we numerically solved the perturbation equation (\ref{eq:perturbations}) by using \Cref{eq:E(a)}, where $w_\phi$ was obtained from \Cref{eq:w} after solving the dynamical system (\ref{system}).
Then, observational constraints on the cosmological parameters were found by maximizing the likelihood probability function of the growth rate factor (GRF) data:
\begin{equation}
\mathcal{L}_\text{GRF}\propto \exp\left[-\dfrac{1}{2}\mathbf{Y}^T \mathcal{C}_{ij}^{-1} \mathbf{Y}\right] ,
\end{equation}
where the vector
\begin{equation}
\mathbf{Y}=\rho(z_i)f\sigma_8^{obs}(z_i)-{f\sigma_8^{th}(z_i)}
\end{equation}
is the difference between the observed values and the values predicted by the theoretical model, once the correction for the fiducial cosmology is taken into account. $\mathcal{C}_{ij}$ is the the covariance matrix constructed by taking into account the correlations between the data points (see \cite{Kazantzidis18} for the details).

We complemented the GRF data with the model-independent observational Hubble data (OHD) obtained through the differential age method \cite{Jimenez02} (see \cite{Abedi18} with the correspondent references). Specifically, the Hubble rate is obtained from the age difference of two nearby, passively evolving, galaxies:
\begin{equation}
H_{th}(z)=-\dfrac{1}{(1+z)}{\left(\dfrac{dt}{dz}\right)}^{-1}\ .
\end{equation}
Since no correlations exist among the OHD measurements, the likelihood function is given as
\begin{equation}
\mathcal{L}_\text{OHD}\propto\exp\left[-\dfrac{1}{2}\displaystyle{\sum_{i=1}^{31}}\left(\dfrac{H_{th}(z_i)-H_{obs}(z_i)}{\sigma_{H,i}}\right)^2\right] \ .
\end{equation}
Since the data samples are independent from each other, the combined likelihood function can be written as
\begin{equation}
\mathcal L_\text{tot}=\mathcal L_\text{GRF}\times \mathcal L_\text{OHD}\ .
\end{equation}
We notice that the growth rate of matter perturbation is insensitive to $H_0$.  As the currently available measurements of the growth rate factor are affected by large uncertainties, we decided to consider in our analysis the Hubble parameter data to be able to obtain tighter constrains on the cosmological parameters. While, from the one hand, this introduces the Hubble constant in the fitting procedure, on the other hand its value would be \textit{de facto} driven by the Hubble parameter data which, by themselves, are not capable of providing a precise estimate of $H_0$. Therefore, keeping $H_0$ as a free parameter in the numerical analysis would only cause a difficulty in constraining the other cosmological parameters on which the present study focuses its attention. For these reasons, we decided to fix $H_0=70$ km/s/Mpc, which represents a safe choice in view of the standing tension between the direct measurements and the model-dependent estimates \cite{Lukovic16,Bernal16}. Moreover, the chosen value is consistent within $1\sigma$ with the results that one obtains from analyzing the Hubble parameter data alone.

Therefore, we constrained the teleparallel dark energy models described in \Cref{sec:dynamics} through Markov Chain Monte Carlo (MCMC) integration method. We implemented the Metropolis algorithm \cite{Metropolis} by means of the software \emph{Mathematica}, sampling over the following parameter space:
\begin{equation}
\mathcal{P}=\{\Omega_{m0},\ \xi,\ \sigma_8\}\ .
\end{equation}
The numerical analysis has been performed assuming the following uniform priors on the fitting parameters:
\begin{equation}
\left\{
\begin{aligned}
&\Omega_{m0}\in(0,1)\ ,\\
&\xi\in(-1,1)\ , \\
&\sigma_{8}\in(0.5,1)\ .
\end{aligned}
\right.
\end{equation}
We ran ten independent chains with 10,000 steps each, starting from a random point of the parameter space. After removing the initial 100 steps of the burn-in phase from each chain, we merged the ten chains to compute the marginalized distributions and the best-fit parameters with the 95\% confidence level.

\begin{figure*}[h!]
\begin{center}
\includegraphics[width=3.2in]{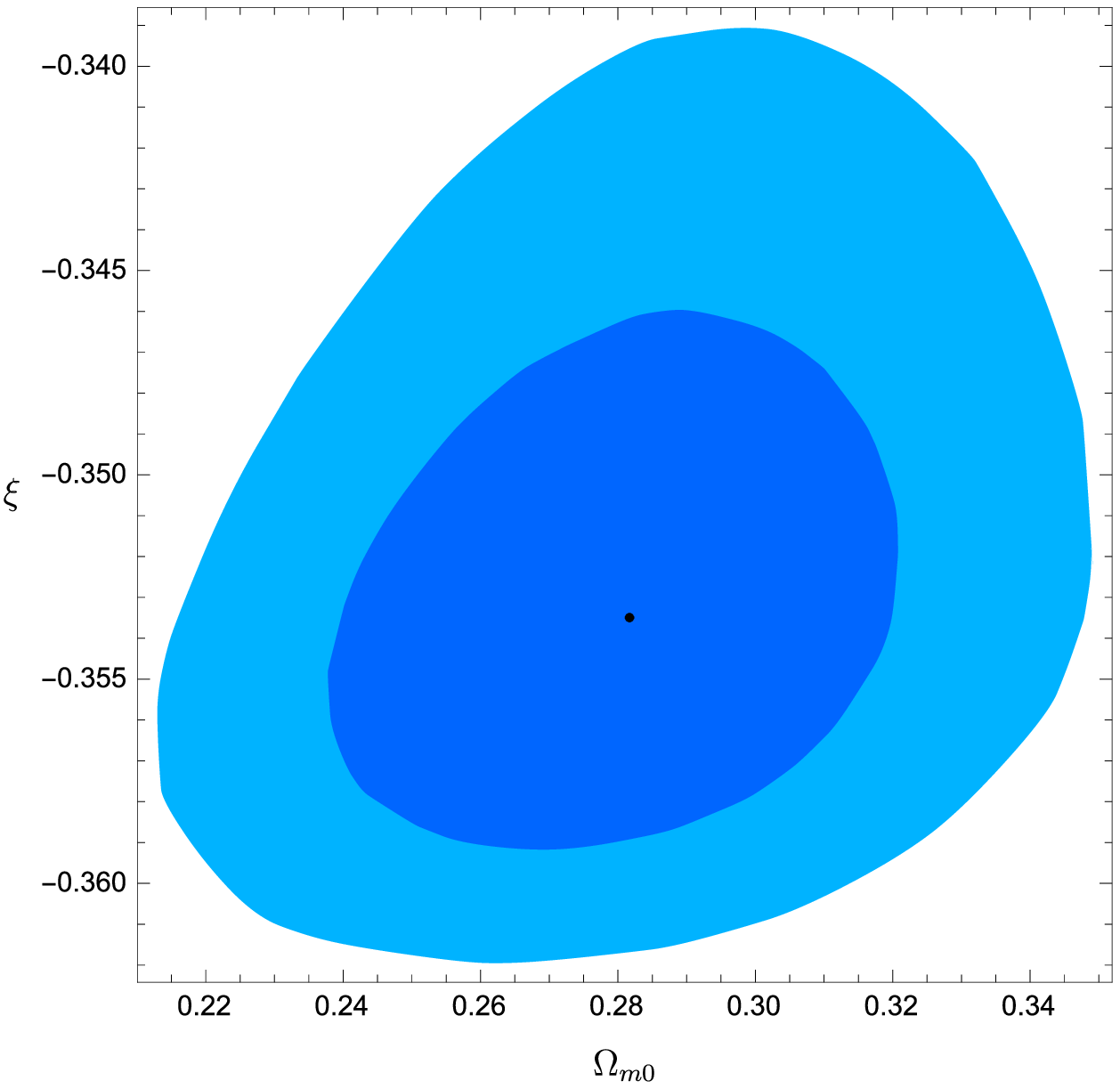}
\hspace{0.5cm}
\includegraphics[width=3.2in]{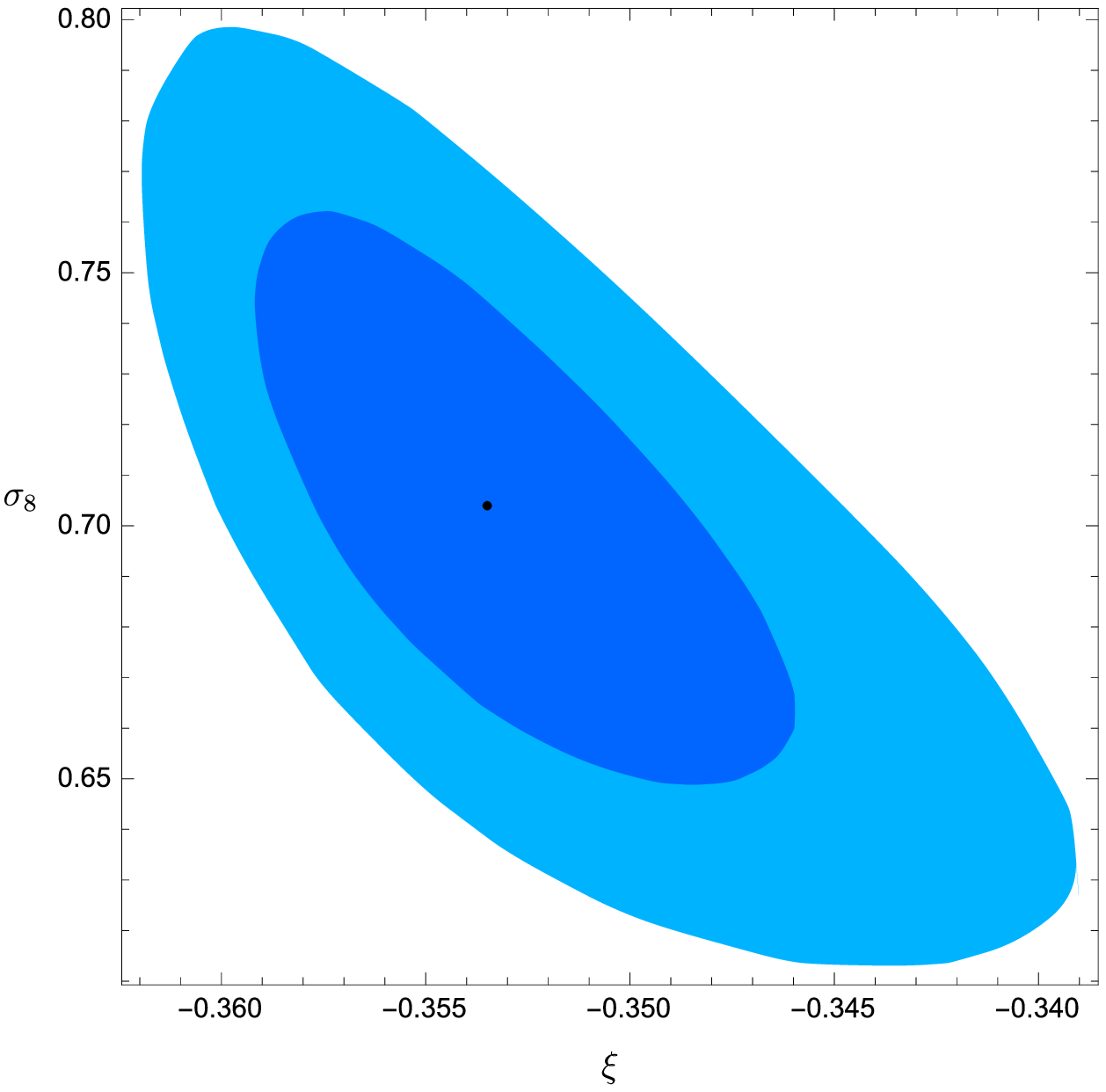}
\caption{Marginalized 68\% and 95\% confidence levels contours resulting from the MCMC analysis on teleparallel dark energy model with $V(\phi)=0$. The black dot corresponds to the best-fit point.}
\label{fig:contours V=0}
\end{center}
\end{figure*}

\begin{figure*}[h!]
\begin{center}
\includegraphics[width=3.2in]{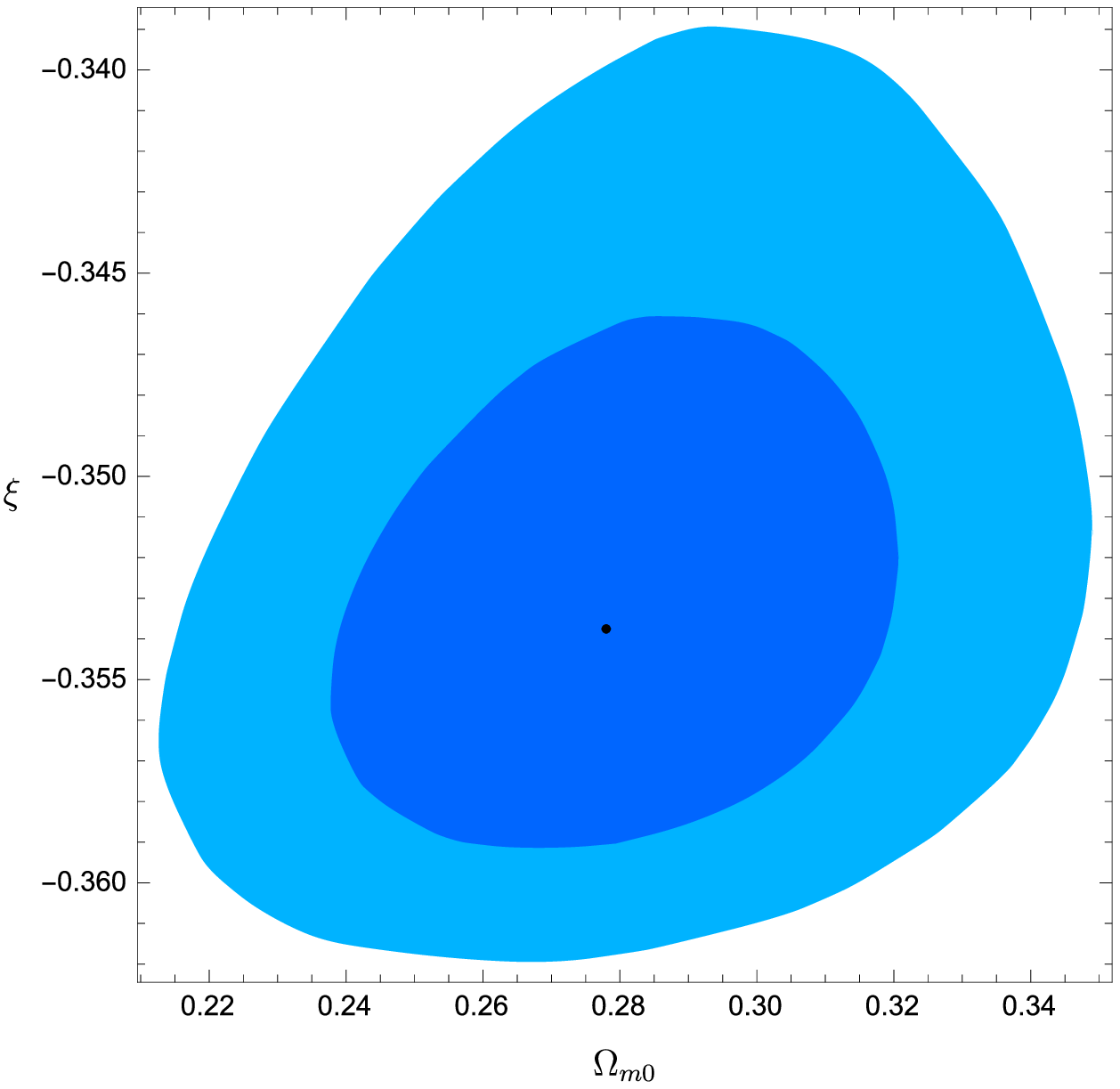}
\hspace{0.5cm}
\includegraphics[width=3.2in]{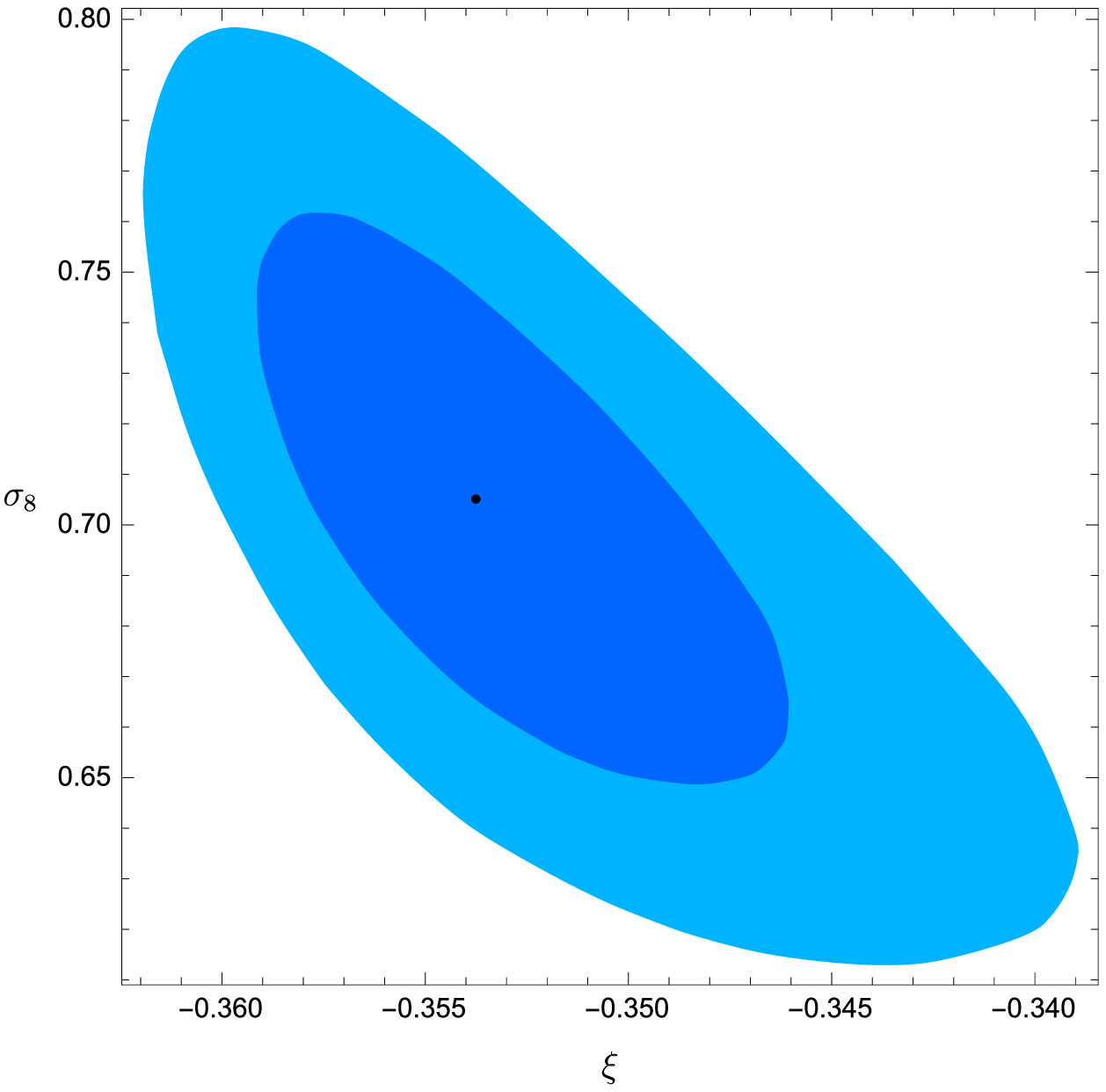}
\caption{Marginalized 68\% and 95\% confidence levels contours resulting from the MCMC analysis on teleparallel dark energy model with $V(\phi)=V_0$. The black dot corresponds to the best-fit point.}
\label{fig:contours V const}
\end{center}
\end{figure*}

\begin{figure*}[h!]
\begin{center}
\includegraphics[width=3.2in]{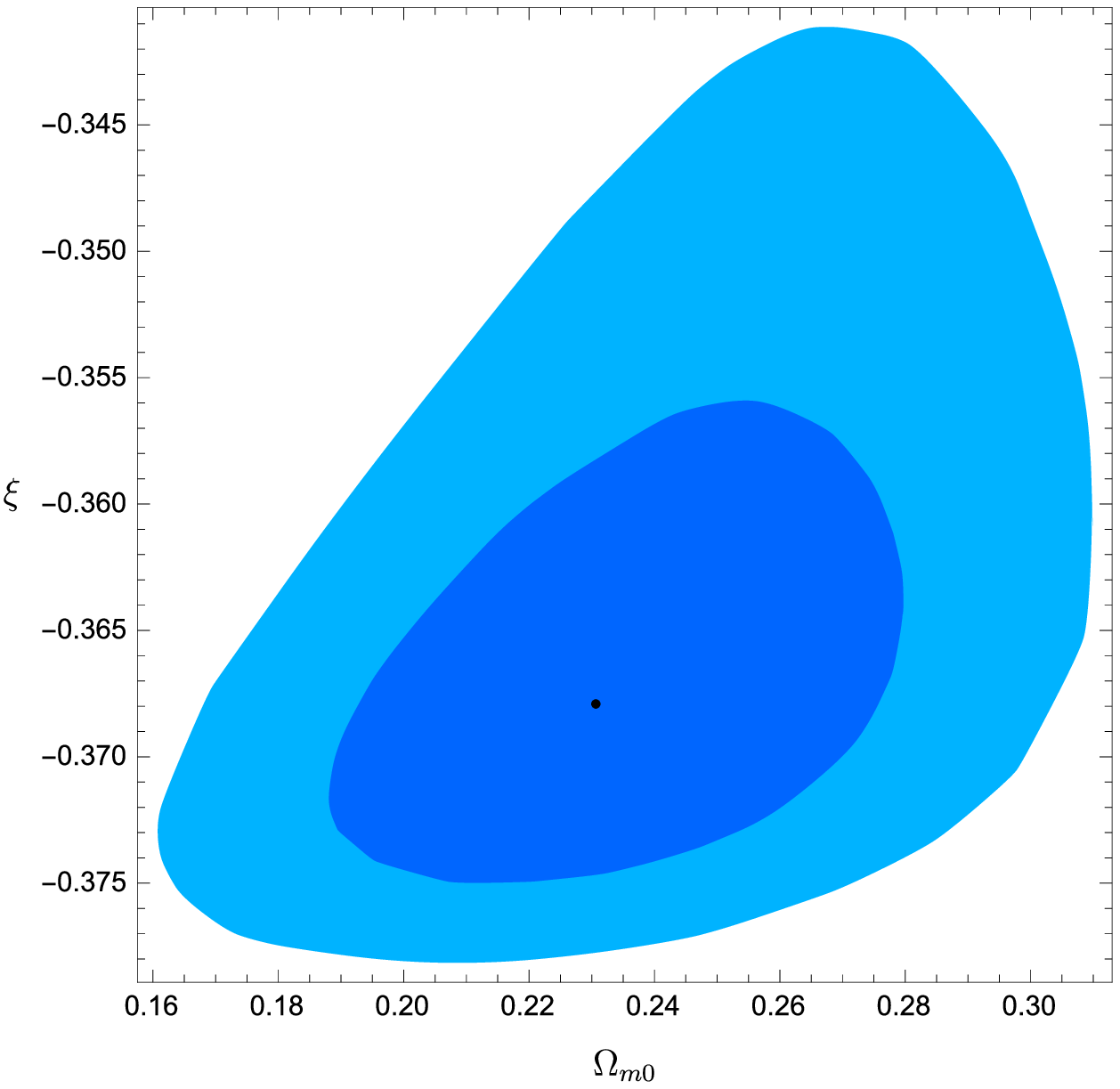}
\hspace{0.5cm}
\includegraphics[width=3.2in]{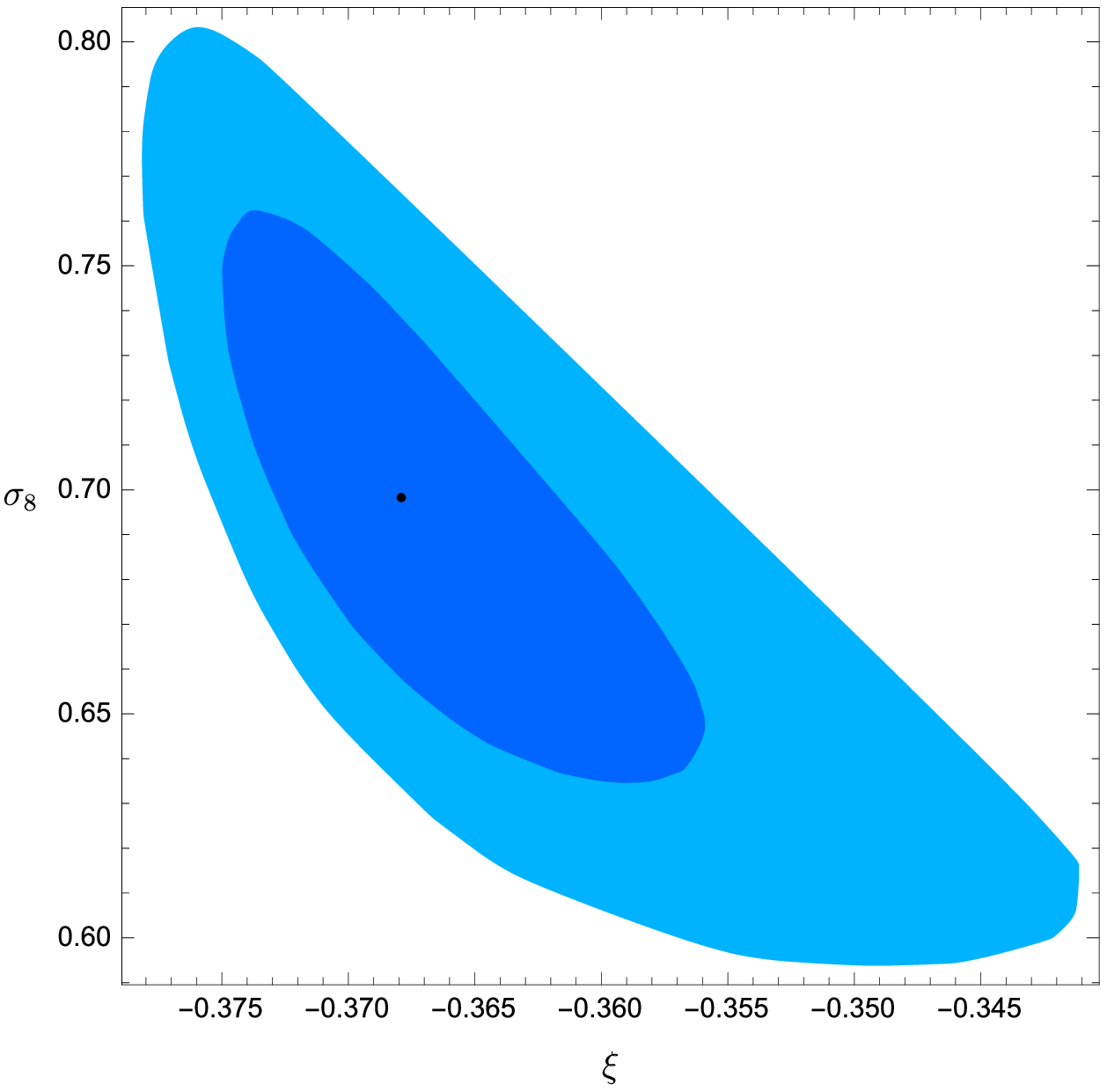}
\caption{Marginalized 68\% and 95\% confidence levels contours resulting from the MCMC analysis on teleparallel dark energy model with $V(\phi)=V_0 \kappa\phi$. The black dot corresponds to the best-fit point.}
\label{fig:contours V linear}
\end{center}
\end{figure*}

\begin{figure*}[h!]
\begin{center}
\includegraphics[width=3.2in]{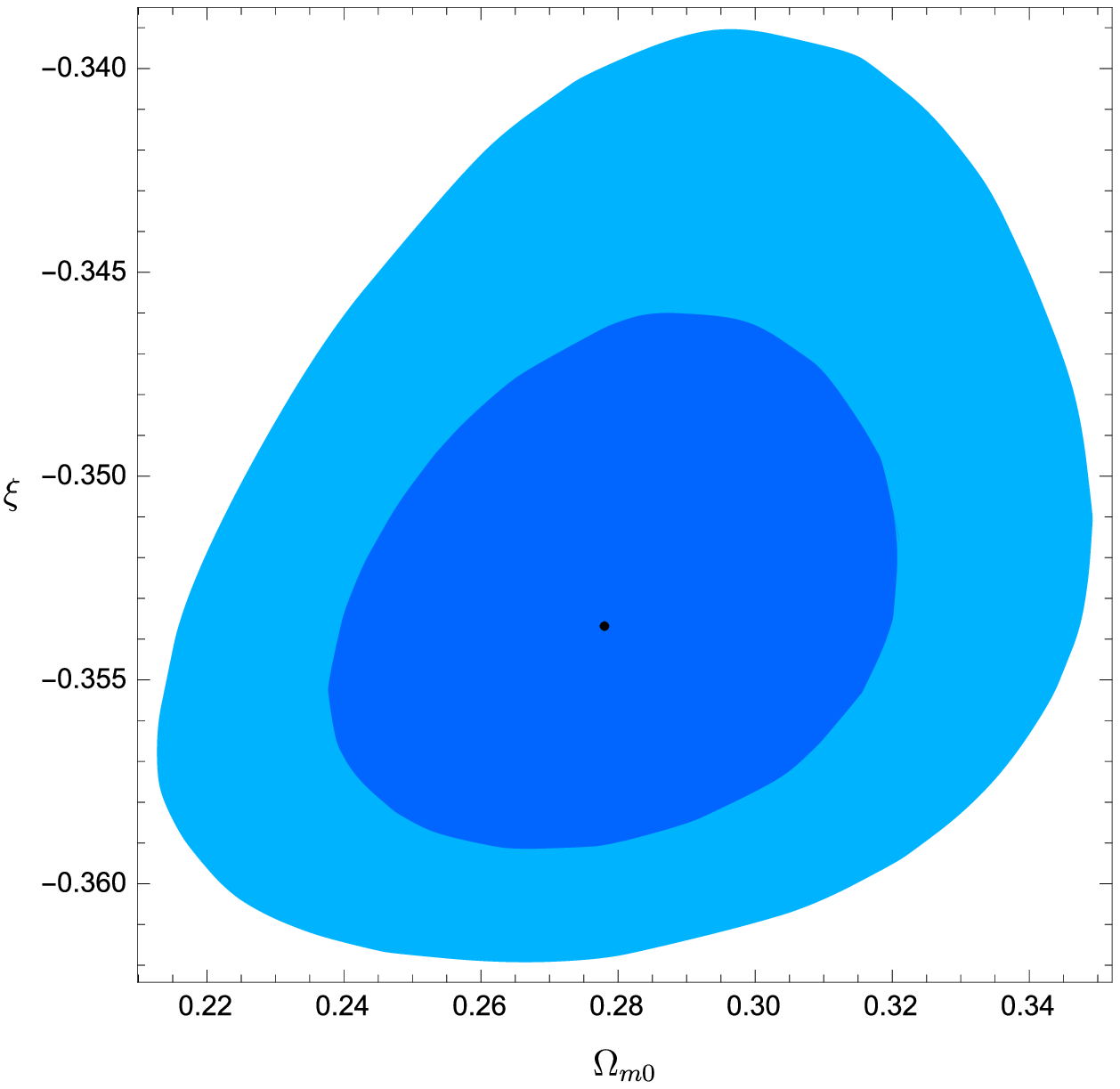}
\hspace{0.5cm}
\includegraphics[width=3.2in]{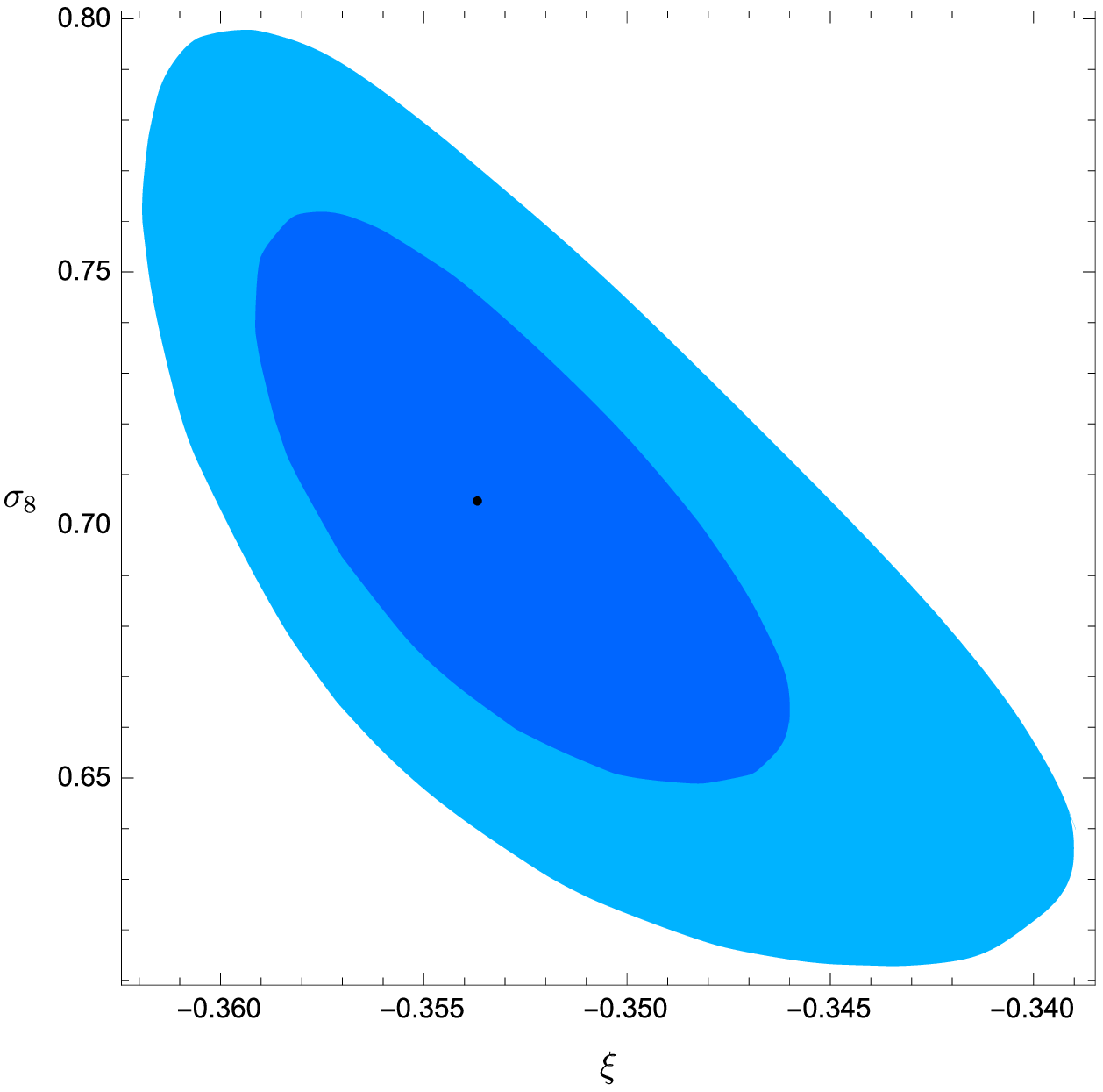}
\caption{Marginalized 68\% and 95\% confidence levels contours resulting from the MCMC analysis on teleparallel dark energy model with $V(\phi)=V_0e^{-\kappa \phi}$. The black dot corresponds to the best-fit point.}
\label{fig:contours V exp}
\end{center}
\end{figure*}

We present in \Cref{tab:results} the best-fit results with relative $1\sigma$ uncertainties for different forms of the scalar field potential.  Furthermore, in Figs.~(\ref{fig:contours V=0})-(\ref{fig:contours V exp}) we show the 2-D $1\sigma$ and $2\sigma$ marginalized likelihood contours for the different models.
We note that all the teleparallel scenarios give similar results, except the model with linear potential.
The cosmological observations clearly constrain the coupling constant to be negative. This result unfortunately implies the non-existence of the attractor solutions found in the analysis of the critical points. However, the obtained values of $\xi$ are consistent with the outcomes got in \cite{Gu13} with other data surveys, with the only exception of the linear potential case.
The $\sigma_8$ values for the teleparallel dark energy models are systematically lower compared to the $\Lambda$CDM prediction, and then in tension with the latest results of the Planck collaboration \cite{Planck18}.
The contours displayed in Figs.~(\ref{fig:contours V=0})-(\ref{fig:contours V exp}) indicate that a non-minimal gravitational coupling scenario is definitely favoured by observations with respect to the minimal quintessence at more than $2\sigma$ evidence.

\clearpage

To compare the different theoretical scenarios, we performed a statistical model selection through the Akaike information criterion (AIC) \cite{Akaike74} and the Bayesian information criterion (BIC) \cite{Schwarz78}. Given a model with $p$ parameters and $N$ data points, the AIC and BIC values are defined as, respectively,
\begin{align}
\text{AIC}=&\ 2p -2\ln \mathcal{L}_\text{max} \ , \\
\text{BIC}=&\ p\ln N -2\ln \mathcal{L}_\text{max}\ ,
\end{align}
where $\mathcal{L}_\text{max}$ is the likelihood value calculated at the best-fit point. The quantities $\Delta(\text{AIC})= \text{AIC}_X-\text{AIC}_Y$ and  $\Delta(\text{BIC})= \text{BIC}_X-\text{BIC}_Y$ provide, between the models $X$ and $Y$, which is the one that better performs against data. This corresponds to the model characterized by the lowest $\text{AIC}$ and $\text{BIC}$ values. In our analysis, we tested the various teleparallel dark energy models against the reference $\Lambda$CDM paradigm. From \Cref{tab:results} one can see that the models with vanishing, constant and exponential potential are statistically disfavoured with respect to the $\Lambda$CDM framework, while the linear potential scenario is ruled out.

\begin{figure}[h!]
\begin{center}
\includegraphics[width=3.in]{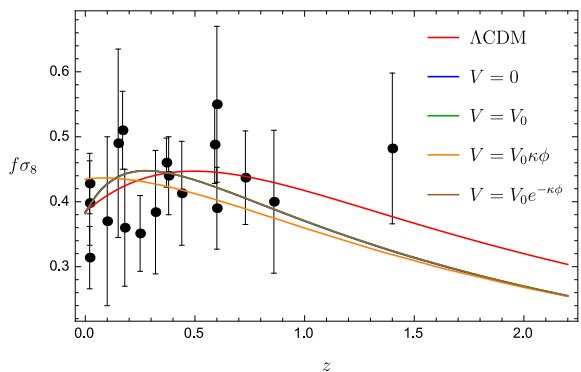}
\caption{Growth rate of matter fluctuations for different teleparallel dark energy scenarios and the $\Lambda$CDM model, assuming the best-fit results obtained from the MCMC analysis. The models with vanishing $(V=0)$, constant $(V=V_0)$ and exponential $(V=V_0e^{-\kappa \phi})$ potential are degenerate.}
\label{fig:comparison}
\end{center}
\end{figure}

\begin{figure}[h!]
\begin{center}
\includegraphics[width=3.in]{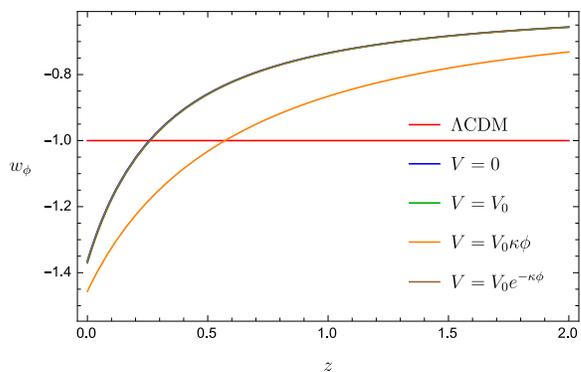}
\caption{Dark energy equation of state parameter for  different teleparallel dark energy scenarios and the $\Lambda$CDM model, assuming the best-fit results obtained from the MCMC analysis. The models with vanishing $(V=0)$, constant $(V=V_0)$ and exponential $(V=V_0e^{-\kappa \phi})$ potential are degenerate.}
\label{fig:w_comparison}
\end{center}
\end{figure}

Finally, adopting the best-fit results obtained from our numerical analysis, we compared in \Cref{fig:comparison} the growth rate for the concordance paradigm and for the different teleparallel models. We notice that the strength of fluctuations at high redshifts is larger for the $\Lambda$CDM model compared to the modified gravity scenarios. With the only exception of linear potential, all the choices for $V(\phi)$ lead to indistinguishable curves which degenerate among them.
Furthermore, in \Cref{fig:w_comparison} we show the behaviour of the dark energy EoS parameter for the different theoretical scenarios. The degeneracy among the models with vanishing, constant and exponential potential cannot be avoided. One can clearly note the characteristic of these models to cross the phantom divide.

\begin{widetext}
\begin{table*}
\begin{center}
\setlength{\tabcolsep}{1.3em}
\renewcommand{\arraystretch}{1.5}
\begin{tabular}{c c c c c c c}
\hline
\hline
Model & $\Omega_{m0}$ & $\xi$  & $\sigma_8$ & $\chi^2_\text{min}$ & $\Delta$AIC & $\Delta$BIC\\
\hline
$\Lambda$CDM & $0.277 \pm 0.024$ & - & $0.785 \pm 0.036$ & 26.9 & 0 & 0 \\
$V(\phi)=0$ & $0.282 \pm 0.027$ & $-0.353 \pm 0.005$ & $0.704 \pm 0.039$ & 31.4 & 6.45  & 8.34 \\
$V(\phi)=V_0$ & $0.278  \pm 0.027 $ & $-0.354 \pm 0.005 $ & $0.705 \pm 0.038$ & 31.4 & 6.45 & 8.34  \\
$V(\phi)=V_0\kappa\phi$ & $0.231 \pm 0.032$ & $-0.368 \pm 0.013 $ & $0.698 \pm 0.047$ & 40.6 & 15.7 & 17.6 \\
$V(\phi)=V_0e^{-\kappa\phi}$ & $0.278  \pm 0.027 $ & $-0.354 \pm 0.005$ & $ 0.705 \pm 0.038 $ & 31.4 & 6.45 &  8.34 \\
\hline
\hline
\end{tabular}
\caption{Best-fit results and $1\sigma$ bounds from the MCMC numerical analysis on the $\Lambda$CDM model and teleparallel dark energy models with different scalar field potentials. The $\chi^2$ values at the best-fit points and the AIC and BIC values with respect to the $\Lambda$CDM model are also shown.}
 \label{tab:results}
\end{center}
\end{table*}
\end{widetext}

\clearpage

\section{Conclusions}
\label{sec:conclusion}

In the present work, we considered modified teleparallel cosmology with a non-minimal coupling between torsion and scalar field.
We showed that teleparallel dark energy exhibits a richer structure of solutions compared to the minimal quintessence. To do that, we described the cosmological evolution in terms of dimensionless variables which form a dynamical system of equations. Assuming suitable forms of the scalar field potential, we were able to solve the autonomous system of equations to find the dynamical behaviour at early and late times. In particular, we analyzed the case of vanishing, constant, linear and exponential potentials. We thus performed a phase-space analysis of the systems and investigated the cosmological solutions in correspondence of the critical points. We also studied the stability of the critical points to search for possible attractor solutions.
Only in the cases of constant and exponential potentials there are late-times attractors, which were found to exist for coupling constant $\xi>0$ and $\xi>1$, respectively.

Then, we studied the growth rate of perturbations. In particular, we wrote the evolution of the matter density contrast, specifying the effective gravitational constant in the quasi-static approximation and in the sub-horizon regime. We thus tested the observational viability of our theoretical scenarios using the growth-rate factor data, which can be considered model-independent after correcting the measurements with respect to the assumed fiducial cosmology. To obtain more stringent constraints on the cosmological parameters, we complemented the growth rate data with the most recent measurements on the Hubble parameter obtained through the model-independent differential age method.  Monte Carlo numerical technique was implemented by means of the Metropolis algorithm assuming uniform distributions for the fitting parameters. The results from the combined likelihood analysis indicate $\xi<0$ at more than $2\sigma$ evidence, implying that a scenario with a non-minimal coupling is strongly favoured over the minimal quintessence scenario.
We found that all the considered models are characterized by amplitudes of matter fluctuations lower than the value predicted by the standard $\Lambda$CDM paradigm.
Moreover, the statistical performance of the various theoretical scenarios was checked through information selection criteria, such as AIC and BIC. Our results show that all the teleparallel dark energy models are disfavoured with respect to $\Lambda$CDM, with a particular strong evidence against the case with linear potential.
Therefore, adopting the best-fit results obtained from our numerical analysis, we computed the dynamical evolution of the dark energy equation of state parameter. We found that the teleparallel models under study can cross the phantom divide, and they are hardly distinguishable from each other with the only exception of the case with linear potential.

For future developments it would be interesting to consider extensions of the present scenario by including a non-zero curvature term. Also, the constraints from the cosmic microwave background anisotropies could play an important role to break the degeneracy among theories with different potentials.

\begin{acknowledgements}
We are grateful to S. Capozziello for useful discussions on the topic of $f(T)$ cosmology. We also want to thank D. Babusci and S. Mancini for their useful suggestions. O. L. thanks the National Institute for Nuclear Physics for financial support and the University of Camerino for its hospitality.
\end{acknowledgements}

\begin{widetext}

\appendix

\section{Coefficients matrix of the perturbation equations for $V(\phi)=V_0$}
\label{app:matrix const}

Here, we report the components of the perturbation coefficients matrix in the case of $V(\phi)=V_0$:

\begin{equation}
\mathcal{M}_{11}=\dfrac{9x^2+8\sqrt{6}\xi ux-3(1+y^2+\xi u^2)}{2(1+ \xi u^2)}\ ,
\end{equation}
\begin{equation}
\mathcal{M}_{12}=-\dfrac{3xy}{1+\xi u^2}\ ,
\end{equation}
\begin{equation}
\mathcal{M}_{13}=-\dfrac{\xi\left(3ux^3-3uxy^2+2\sqrt{6}x^2(-1+u^2\xi)+\sqrt{6}(1+\xi u^2)^2\right)}{(1+\xi u^2)^2}\ , \\
\end{equation}
\begin{equation}
\mathcal{M}_{21}=\dfrac{y (3x  + 2 \sqrt{6} u\xi)}{1 + \xi u^2}\ , \\
\end{equation}
\begin{equation}
\mathcal{M}_{22}=\dfrac{3 (1 + x^2 - 3 y^2) + 3\xi u^2 +  4 \sqrt{6}\xi x u}{2 (1 + \xi u^2)}\ , \\
\end{equation}
\begin{equation}
\mathcal{M}_{23}=-\dfrac{y\xi \left(3u(x^2-y^2)+2\sqrt{6}x(-1+\xi u^2)\right)}{2u^2(1+\xi u^2)^2}\ , \\
\end{equation}
\begin{equation}
\mathcal{M}_{31}=\sqrt{6}\ , \\
\end{equation}
\begin{equation}
\mathcal{M}_{32}=\mathcal{M}_{33}=0\ .
\end{equation}

\section{Coefficients matrix of the perturbation equations for $V(\phi)=V_0\kappa\phi$}
\label{app:matrix lin}

Here, we report the components of the perturbation coefficients matrix in the case of $V(\phi)=V_0\kappa\phi$:

\begin{align}
&\mathcal{M}_{11}=\dfrac{9x^2+8\sqrt{6}\xi ux-3(1+y^2+\xi u^2)}{2(1+\xi u^2)}\ , \\
&\mathcal{M}_{12}=-\dfrac{\sqrt{6}y}{u}-\dfrac{3xy}{1+\xi u^2}\ , \\
&\mathcal{M}_{13}=\dfrac{y^2(\sqrt{6}+2\sqrt{6}\xi u^2+6\xi x u^3+\sqrt{6}\xi^2 u^4)-2\xi u^2\left(3x^3 u+2\sqrt{6}x^2(-1+\xi u^2)+\sqrt{6}(1+\xi u^2)^2\right)}{2u^2(1+\xi u^2)^2}\, \\
&\mathcal{M}_{21}=\dfrac{y (\sqrt{6} + 6 x u + 5 \sqrt{6}\xi u^2)}{2 u(1 + \xi u^2)}\ , \\
&\mathcal{M}_{22}=\dfrac{\sqrt{6} x + 3 u (1 + x^2 - 3 y^2) + 3\xi u^3 +  5 \sqrt{6}\xi x u^2}{2 u(1 + \xi u^2)}\ , \\
&\mathcal{M}_{23}=\dfrac{y\left(-6 \xi x^2 u^3 + 6\xi y^2 u^3  - \sqrt{6} x (1 - 2\xi u^2+ 5 \xi^2 u^4)\right)}{2u^2(1+\xi u^2)^2}\ , \\
&\mathcal{M}_{31}=\sqrt{6}\ , \\
&\mathcal{M}_{32}=\mathcal{M}_{33}=0\ .
\end{align}

\section{Coefficients matrix of the perturbation equations for $V(\phi)=V_0e^{-\kappa\phi}$}
\label{app:matrix exp}

Here, we report the components of the perturbation coefficients matrix in the case of $V(\phi)=V_0e^{-\kappa\phi}$:

\begin{align}
&\mathcal{M}_{11}=\dfrac{9 x^2 + 8 \sqrt{6}\xi x u - 3 (1 + y^2 + \xi u^2)}{2(1+\xi u^2)}\ , \\
&\mathcal{M}_{12}=y \left(\sqrt{6} - \dfrac{3x}{1+\xi u^2}\right) , \\
& \mathcal{M}_{13}=-\dfrac{\xi \left(3 x^3 u - 3 x y^2 u + 2 \sqrt{6} x^2 (-1 + \xi u^2 ) + \sqrt{6} (1 + \xi u^2)^2\right)}{(1 + \xi u^2)^2}\ , \\
& \mathcal{M}_{21}=-\dfrac{y \left(-6 x + \sqrt{6} (1 - 4\xi u + \xi u^2)\right)}{2(1+\xi u^2)}\ , \\
& \mathcal{M}_{22}= \dfrac{3(1 + x^2 - 3 y^2 + \xi u^2)  - \sqrt{6} x (1 - 4\xi u + \xi u^2)}{2(1+\xi u^2)}\ , \\
& \mathcal{M}_{23}=-\dfrac{\xi y \left(3u(x^2 + y^2) + 2 \sqrt{6} x (-1 + \xi u^2)\right)}{(1 +\xi u^2)^2}\ , \\
&\mathcal{M}_{31}=\sqrt{6}\ , \\
&\mathcal{M}_{32}=\mathcal{M}_{33}=0\ .
\end{align}

\end{widetext}

\end{document}